# First observation of the solar Type III burst decay and its interpretation

V.N. Melnik (1), A.A. Konovalenko (1), S.M. Yerin (1), I.M. Bubnov (1), A.I. Brazhenko (2), A.V. Frantsuzenko (2), V.V. Dorovskyy (1), M.V. Shevchuk (1), H.O. Rucker (3)

(1) Institute of Radio Astronomy, National Academy of Sciences of Ukraine, Kharkov, Ukraine

(2) Gravimetrical Observatory, Poltava, Ukraine

(3) Commission for Astronomy, Austrian Academy of Sciences, Graz, Austria

**Abstract**

A decay of Type III burst into two Type III bursts was registered during solar observations by GURT and URAN-2 radio telescopes on April 18, 2017. Such phenomenon was observed for the first time. Newborn Type III bursts have drift rates smaller than that of decaying Type III burst. Such decays of Type III bursts were predicted by a gas-dynamic theory of high-energy electron beams propagating through the thermal background plasma. In the frame of this theory Type III sources are beam-plasma structures moving with constant velocities. In our case the sum of velocities of newborn Type III sources equals the velocity of decaying Type III source. The last one is $0.33c$ in the case of fundamental radio emission and $0.2c$ at the harmonic radio emission of Type III burst. The density ratio of slow and fast newborn Type III sources is about 3.

## 1. Introduction

Type III bursts are observed during more than 60 years. Wild (1950) was the first, who registered and described their properties. These bursts are tracks of radio emission, which drift from high frequencies to low frequencies. Profiles of Type III bursts have a quick rise and slower falling. Such form is registered at every frequency from 1 GHz to tens kHz (Suzuki and Dulk, 1985). Type III durations decrease with frequency.

Already in the first paper (Wild, 1950) Wild supposed that electrons with high velocities, up to $0.3c$ (c is the speed of light), were responsible for Type III bursts based on their high frequency drift rates.

Ginzburg and Zhelezniakov (1958) suggested plasma mechanism of Type III bursts radio emission. This mechanism consisted in generation of Langmuir waves by fast electrons due to two-stream instability and their subsequent transformation into electromagnetic waves in processes of Langmuir waves scattering on ions (fundamental) or Langmuir waves coalescence (second harmonic). Later (Melrose, 1985) an interaction of Langmuir waves with ion sound waves was attracted for generation of the fundamental Type III bursts. The key point of plasma mechanism is generation of Langmuir waves. These waves are both generated and absorbed by fast electrons at every point. These processes lead to a formation of fast electron distribution function. A plateau on the distribution function from small velocities to the maximum velocity establishes very quickly (Drummond, and Pines, 1962, Vedenov et al., 1962), during the quasilinear time $\tau_{qu} = (\omega_{pe} n'/n)^{-1}$ ($\omega_{pe} = \sqrt{4\pi e^2 n/m}$ is the plasma frequency, $n'$ is the fast electron density, and $n$ is the plasma density, $e$ and $m$ are the electron charge and mass). After that the generation of Langmuir waves stops and Type III radio emission ceases. As a result Type III bursts must exist only in the narrow frequency band near the place of electron injection. At the same time observations show that Type III bursts are registered in a very wide frequency band. All this formulates so-called Sturrok's dilemma (Sturrock, 1964). Its resolution is in the fact that electron beams are space limited (Zheleznyakov and Zaitsev, 1970, Melrose, 1985). This leads to the fastest electrons of beam forehand come to some point with a distribution function with a positive derivative and as a consequence these electrons generate Langmuir waves in this point. At the following moments, when slower electrons come to this point, they absorb some Langmuir waves and increase their velocities. Consequently a nonlinear plasma object, a beam-plasma structure, is formed (Mel'Nik, 1995, Mel'nik, et al., 1999a). It consists of electrons and Langmuir waves and can propagate with a constant velocity at large distances. Langmuir waves of beam-plasma structure are transformed into electromagnetic waves at the local plasma frequency and double plasma frequency giving two harmonics of radio emission according to Ginzburg and Zhelezniakov (1958). Brightness temperatures of two harmonics can explain those of Type III bursts (Mel'Nik, and Kontar, 2003a, Ratcliffe et al., 2014). Moreover the explanation of the fact, that Type III sources have velocities of about 0.3c, can be understood too (Mel'Nik, 2003b).

There are different developments of electrons flying off through plasma in the dependence of initial electron distribution functions and conditions in the coronal plasma. If the initial function is a monoenergetic one then the beam-plasma structure with the constant velocity equaled half of initial electron velocity propagates through plasma and there is a high level of Langmuir waves in the place of electron injection (Mel'Nik, 1995, Mel'nik, et al., 1999a). Inhomogenuities of coronal plasma lead to decelerating beam-plasma structures and as a consequence to decreasing frequency drift rates of Type III bursts (Kontar, 2001a). Coronal plasma with density fluctuations changes the spatial distribution of generated Langmuir waves (Kontar, 2001b) that can be manifested as Type IIIb bursts (Kontar et al., 2017). Influence of electron beam densities, their hardness and temperature of coronal plasma on properties of beam-plasma structures were recently studied by Reid and Kontar (2018). In the case of an external electrical field a beam-plasma structure moves with acceleration (Kontar, and Melnik, 2003). If the initial electron distribution function is a Maxwellian one, then a part of electrons are locked by Langmuir turbulence near the place of injection and a small part of fastest electrons propagate into plasma as a beam-plasma structure (Melnik, et al., 2000).

In the case of injection of two monoenergetic electron beams there are some scenarios in dependence of densities and velocities of these beams (Melnik and Kontar, 1998a, Melnik and Kontar, 1998b). One of them is the formation of beam-plasma structure, which decays into two beam-plasma structures at some distance. In application to Type III bursts a decay of Type III burst into two Type III bursts (Melnik and Kontar, 1997) should be observed.

In this paper we report the registration of such case in solar observations in the wide frequency band of 8-70 MHz by GURT and URAN-2 radio telescopes on April 18, 2017.

## 2. Observations

We used two radio telescopes during the observation campaign, a subarray of GURT radio telescope (8-80 MHz) and URAN-2 radio telescope (8-33 MHz) (Konovalenko et al. 2016).

URAN-2 radio telescope is a regular antenna array of 512 elements (Brazhenko et al., 2005). Each element is a combination of two perpendicular fat dipoles. Signals of each linear polarization are amplified, phased and combined in analog way. The estimated effective area at 20 MHz is 28 000 m$^2$. DSPZ digital receiver (Ryabov et al., 2010, Zakharenko et a;., 2016) is used for digitization, spectra computations and data storage. During the solar observational campaign the receiver recorded data with 4 kHz-100 ms frequency-time resolution. URAN-2 radio telescope can measure polarization of registered radio emission.

The GURT radio telescope situated near the UTR-2 radio telescope is a new instrument in progress. Now only one sub-array is used for solar observations. The square sub-array consists of 25 active elements (Konovalenko et al. 2016). The distance between elements is 3.75 m. The effective area of the subarray is about 350 m$^2$ at 40 MHz, so its sensitivity is worse than that for the URAN-2 radio telescope in the frequency band 8-33 MHz. An ADR digital receiver (Zakharenko et al., 2016) was used for registration. The frequency-time resolution during solar observations was 9 kHz-100 ms.

Some solar radio phenomena in the decameter range were observed during a CME on April 18, 2017. There were Type IV burst, two groups of Type III bursts preceding it, Type II burst with herring-born structure, inverted U-bursts with developed ascendant and descending legs (Figure 1) and decaying Type III burst (Figure 2) discussed in this paper. All of these phenomena were associated with the active region NOAA 2651, which was situated at once behind the eastern limb (Figure 3). STEREO A shows that a flare happened in this region in the time interval 9:26-9:36 UT (Figure 3). Maximum of X-ray emission with energy <25 keV was approximately at 9:45 UT so we can conclude that the discussed Type III burst (Figure 2) occurred at the increasing phase of X-ray emission (at 9:34:25 UT at 70 MHz) and it seems that electrons responsible for this Type III burst have been accelerated in the flare.

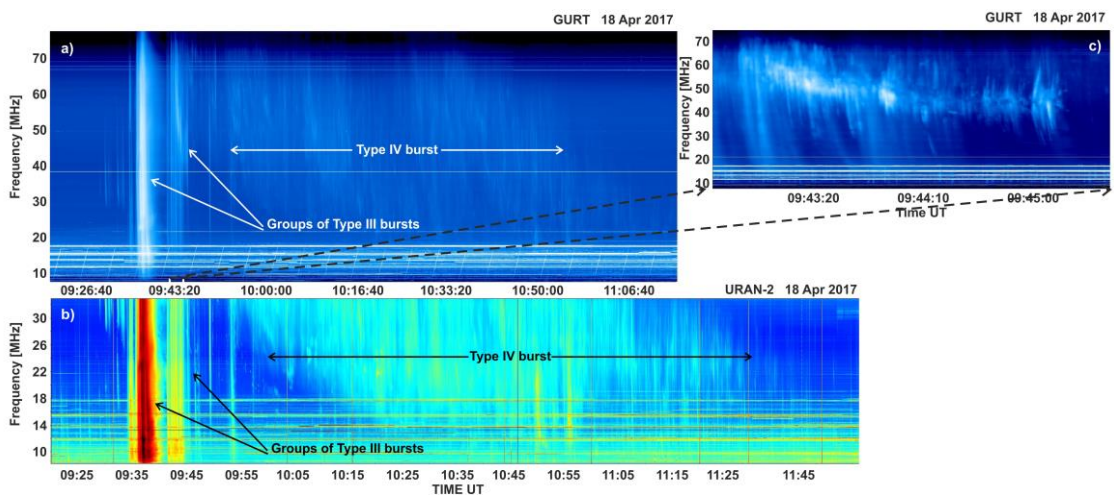

**Figure 1**. Type IV burst, groups of Type III bursts (a,b), and Type II burst (c) according to observations by GURT and URAN-2 radio telescopes.

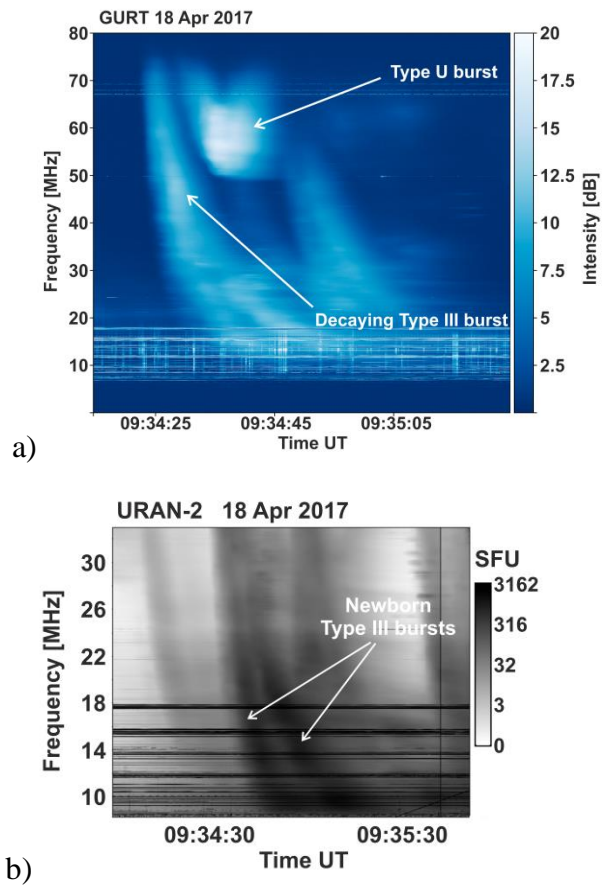

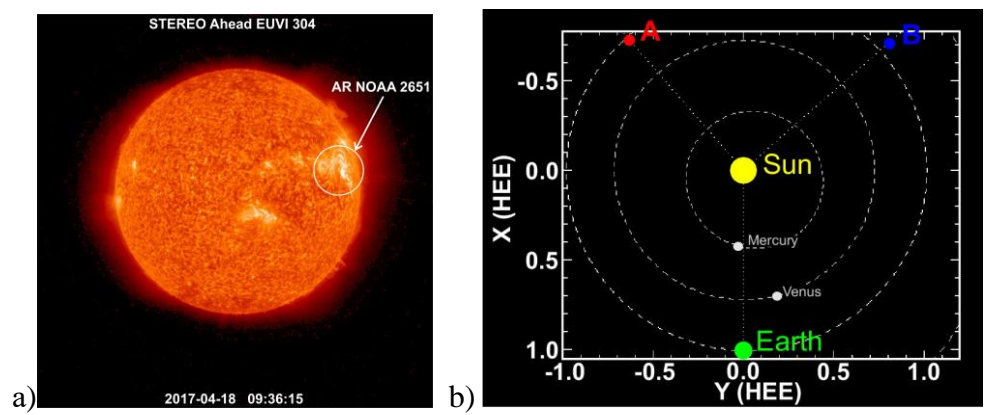

**Figure 2.** Type III bursts in the frequency bands of 8-80 MHz ( GURT ) (a) and 8-33 MHz (URAN-2) (b).

**Figure 3.** The active region NOAA 2651 observed by STEREO A (a) and the position of STEREO A (b) on April 18, 2017.

Profiles of decaying Type III burst and newborn Type III bursts are presented in the Figure 4. We see that there Type III burst drifted from high to low frequencies with a drift rate of about 4.4 MHz s$^{-1}$ at 32 MHz and it decayed into new two Type III bursts in the frequency region of 30-35 MHz (Figure 4, 5). These newborn Type III bursts drifted from high to low frequencies as well but with smaller drift rates, 2.7 and 1.7 MHz s$^{-1}$ respectively, at frequency 32 MHz (Figure 5b). Data from the radio telescope URAN-2 are shown in the Figures 3, 4 and 5 in the frequency range 8-33 MHz because its sensitivity in this range is essentially better than the sensitivity of GURT and we can follow the newborn Type III bursts down to 9 MHz.

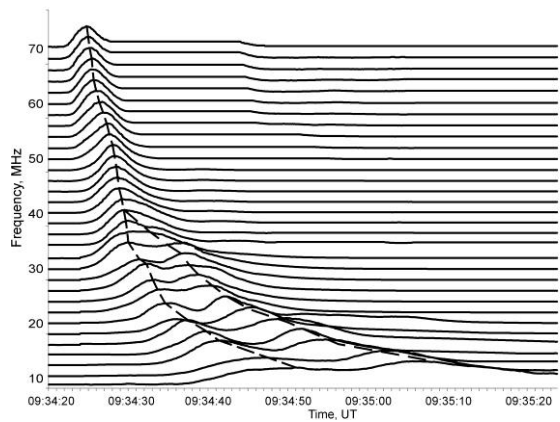

**Figure 4.** Profiles of decaying Type III burst and two newborn Type III bursts according to GURT (32-72 MHz) and URAN-2 (8-32 MHz) data. Each profile is normalized by a maximum flux at a given frequency.

Their radio fluxes increased from 50 s.f.u. at 30 MHz to 1400 s.f.u. at 9 MHz. The full width half maximum (FWHM) duration of decaying Type III burst was 8.3 s at 35 MHz and durations of newborn Type III bursts at level of 90% were 3-3.6 s at 30 MHz. According to URAN-2 data polarizations of these bursts were not higher than 5%. It says that Type III bursts were probably harmonic (Suzuki and Dulk, 1985, Melnik, et al., 2018). The discussed Type III burst was generated by electrons accelerated in the flare situated at once behind the eastern limb. So these electrons moved practically in the picture plane. The directional pattern of the fundamental radio emission is narrow and it is directed in the direction of electrons propagation (Zheleznyakov, 1977) and thus this radio emission can be hardly registered on the Earth. At the same time, the second harmonic of the radio emission has a wide directional pattern (Zheleznyakov, 1977) and can easily be observed by the ground-based radio telescopes. Recently some Type III bursts associated with electrons propagating in the picture plane were registered in interferometer observations by the UTR-2 radio telescope (Melnik, et al., 2017) and it was demonstrated that those Type III bursts were the second harmonics. All this supports our assumption that the observed Type III bursts are harmonics.

At the same time as was shown in (Sirenko et al, 2002) the nonlinear interaction of O-mode and X-mode radio waves with short wavelength kinetic Alfven waves (KAWs) can depolarize initial radio emission at a high level of KAWs turbulence (Lyubchyk et al., 2017) and we shall consider the assumption that observed bursts were fundamentals, too.

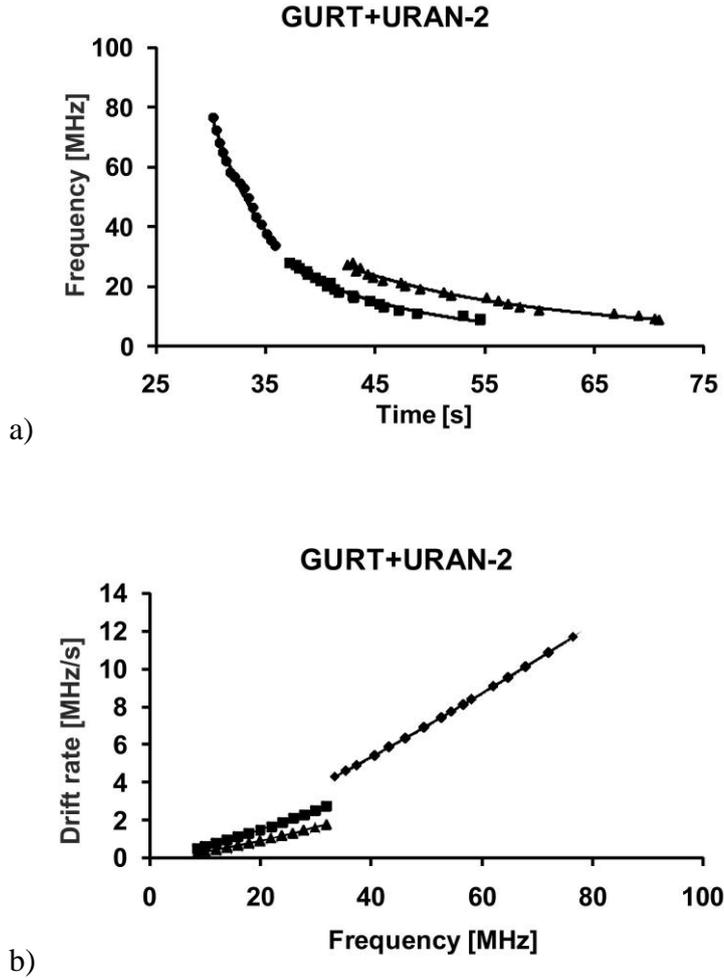

a)

b)

**Figure 5.** Tracks of flux maximums of Type III bursts (a) and their drift rates on frequency (b) according to GURT (32 -75 MHz) and URAN-2 (9-32 MHz) data.

### 3. Discussion

Melnik (1995) proposed a gas-dynamic model of electron beam propagation through plasma, when an influence of generation and absorption of Langmuir waves on propagation was essential. In such nonlinear situation a beam-plasma structure consisting of electrons and Langmuir waves is formed (Melnik, 1995, Melnik et al.,1999b) and reminds a soliton (Melnik, et al., 1999a). This beam-plasma structure moves with a constant velocity at large distances without energy loss. Moreover, because it consists of Langmuir waves it can be a source of radio emission and a source of Type III burst, when such a structure moves through plasma of solar corona (Melnik, 2003, Ratcliffe, et al., 2014). Gas-dynamic equations were obtained from kinetic quasilinear equations of weak turbulence (Vedenov, et al., 1962, Drummond and Pines, 1962) for electron distribution functions $f(v,x,t)$ and spectral energy density of Langmuir waves $W(k,x,t)$

$$\frac{\partial f}{\partial t} + v\frac{\partial f}{\partial x} = \frac{4\pi^2 e^2}{m^2}\frac{\partial}{\partial v}\frac{W}{v}\frac{\partial f}{\partial v} \qquad (1)$$

$$\frac{\partial W}{\partial t} = \frac{\pi \omega_{pe}}{n} v^2 W \frac{\partial f}{\partial v}, \qquad \omega_{pe} = kv$$

supposing that processes of interactions of electrons and Langmuir waves (quasilinear time $\tau_{qu} = (\omega_{pe} \frac{n'}{n})^{-1}$) happen quickly in comparison with the flying off time $t$. The situation is analogous to that which arises at the transition from kinetic equations to usual gas-dynamic equations for a gas (Silin, 1971) at quick collisions between gas particles. In our case gas-dynamic equations are formulated for the plateau height $p(x,t)$ and the maximum velocity $u(x,t)$ of the plateau because a plateau is established (Melnik, 1995, Melnik, et al., 1999b)

$$f(v,x,t) = \begin{cases} p(x,t), & v < u(x,t) \\ 0, & v > u(x,t) \end{cases} \qquad (2)$$

during the quick quasi-linear time.

The spectral energy density of Langmuir waves has the form

$$W(v,x,t) = \begin{cases} W_0(v,x,t), & v < u(x,t) \\ 0, & v > u(x,t) \end{cases} \qquad (3)$$

Gas-dynamic equations are as following

$$\frac{\partial}{\partial t} pu + \frac{\partial}{\partial x} \frac{pu^2}{2} = 0$$

$$\frac{\partial u}{\partial t} + u \frac{\partial u}{\partial x} = 0 \qquad (4)$$

$$\frac{\partial p}{\partial t} + v \frac{\partial p}{\partial x} = \frac{\omega_{pe}}{m} \frac{\partial}{\partial v} \frac{1}{v^3} \frac{\partial W_0}{\partial t}$$

with boundary conditions for $W_0(k,x,t)$

$$\frac{\partial W_0}{\partial t} = 0, \qquad v = u$$

$$\frac{\partial u}{\partial t} W_0 = 0, \qquad v = u \qquad (5)$$

In the case of a monoenergetic electron beam $f(v) = f_0 \delta(v - v_0)$ solution of equations (4-5) is a beam plasma structure

$$u(x,t) = const = v_0$$

$$p(x,t) = p(x - \frac{v_0 t}{2}) \qquad (6)$$

$$W_{0,pl} = \frac{m}{\omega_{pe}} v^4 (1 - \frac{v}{v_0}) p(x - \frac{v_0 t}{2}),$$

8which propagates with the constant velocity $v_{pl} = v_0/2$ (Melnik, 1995, Melnik, et al., 1999b).

In the case, when the boundary electron distribution function splits into two monoenergetic beams with velocities $u_1$ and $u_2 > u_1$

$$f(v, x=0, t) = [n_1 \delta(v - u_1) + n_2 \delta(v - u_2)] \exp(-t/\tau) \tag{7}$$

at the condition

$$\frac{n_1}{u_1} > \frac{n_2}{u_2 - u_1}, \tag{8}$$

solution of corresponding gas-dynamic equations is a beam-plasma structure with electrons with the distribution function consisting of two plateaus

$$f(v, x, t) = \begin{cases} p_1(x,t), & v < u_1 \\ p_2(x,t), & u_1 < v < u_2 \\ 0, & v > u_2 \end{cases} \tag{9}$$

and spectral energy density for Langmuir waves

$$W_s(v, x, t) = \begin{cases} W_1(x,t), & v < u_1 \\ W_2(x,t), & u_1 < v < u_2 \\ 0, & v > u_2 \end{cases} \tag{10}$$

It moves with velocity $v_{decay} = (u_1 + u_2)/2$ (Melnik, and Kontar, 1998b) up to the distance

$$x^* = \frac{(u_1 + u_2)(2u_2 - u_1)\tau}{2u_2} \ln\left(\frac{n_1(u_2 - u_1)}{n_2 u_1}\right) \tag{11}$$

from the place of injection. At larger distances $x > x^*$ the solution consists of two beam-plasma structures (6) moving with velocities $v_1 = u_1/2$ and $v_2 = u_2/2$ and as we see the sum of these velocities equals the velocity of decaying beam-plasma structure $v_{decay} = v_1 + v_2$. As far as each beam-plasma structure is a source of radio emission in the form of Type III burst then decay of Type III burst into two Type III bursts with smaller drift rates must be observed at height $x > x^*$.

In our case taking into account that newborn Type III bursts were generated at the second harmonic in the Newkirk corona (Newkirk, 1961), we derived velocities of beam-plasma structures $v_1 = 0.4 \cdot 10^{10} cms^{-1}$ and $v_2 = 0.63 \cdot 10^{10} cms^{-1}$ from the equation for the frequency drift rate

$$df/dt = f/2 \cdot dn/ndr \cdot v \tag{12}$$

and values of frequency drift rates for newborn Type III bursts 1.7 and 2.7 MHz s$^{-1}$. The beam-plasma structure of decaying Type III burst had velocity $v_{decay} = (v_1 + v_2) = 1.03 \cdot 10^{10} cms^{-1}$. This value is equal to the velocity derived from the frequency drift rate of this burst in the Newkirk model (Newkirk, 1961) at 32 MHz.

According to equation (11) the distance, at which Type III burst decay happened, equaled $R = 2.3 R_S$ in the Newkirk model (from this distance radio emission of the second harmonic at 32 MHz is released) if the density ratio of slow and fast beam-plasma structures was $n_1/n_2 \approx 3$. This ratio looks quite reasonable because fluxes of both newborn Type III bursts, which are defined by densities and velocities of their sources, are comparable.

If the discussed Type III bursts were fundamentals then radio emission escaped from the coronal plasma at the distance $R = 1.74 R_S$ and velocities of beam-plasma structures responsible for newborn Type III bursts were $v_{decay} = 0.59 \cdot 10^{10} cms^{-1}$, $v_1 = 0.23 \cdot 10^{10} cms^{-1}$ and $v_2 = 0.36 \cdot 10^{10} cms^{-1}$, which were approximately two times smaller than in the case, if radio emissions were the second harmonics. In both cases, if Type III bursts were fundamentals or harmonics, velocities of Type III sources seem to be rational. Note that the density ratio of slow and fast beam-plasma structures is approximately the same $n_1/n_2 \approx 3$ in both cases, i. e. this ratio does not essentially depend on velocities of beam-plasma structures.

Thus we see that the main properties of decaying and new born Type III bursts can be understood in the frame of gas-dynamic theory of flying off of fast electrons passing through the coronal plasma.

## 4. Summary

We report about the first observation of Type III burst decaying into two separate Type III bursts. Observations were carried out in the wide frequency band from 8 to 80 MHz with URAN-2 and GURT radio telescopes on 18 April 2017. Decaying Type III burst drifted from a frequency of 80 MHz towards lower frequencies and decayed approximately at 32 MHz into two Type III bursts, which then drifted slower down to 9 MHz. Type III bursts are radio emissions of beam-plasma structures moving with constant velocities in the solar corona in the gas-dynamic theory of fast electrons propagation through plasma. According to this theory such beam-plasma structure can split into two independent beam-plasma structures with smaller velocities. In the dynamic spectrum the emission looks like a decay of Type III burst into two newborn Type III bursts with smaller drift rates. In our case the velocity of beam-plasma structure responsible for decaying Type III burst was about $10^{10} cms^{-1} = 0.33c$ (if decaying Type III burst was harmonic) and $0.6 \cdot 10^{10} cms^{-1} = 0.2c$ (if the decaying Type III burst was fundamental) and velocities $0.4 \cdot 10^{10} cms^{-1}$ and $0.59 \cdot 10^{10} cms^{-1}$ for harmonics and $0.23 \cdot 10^{10} cms^{-1}$ and $0.36 \cdot 10^{10} cms^{-1}$ for fundamentals for beam-plasma structures of newborn Type III bursts. The density ratio of slow and fast beam-plasma structures was about 3 in both cases.

The research was supported by projects "Radio telescope" and "Spectr-3" of National Academy of Sciences of Ukraine.